\documentclass{sigchi-ext}
\usepackage[T1]{fontenc}
\usepackage{textcomp}
\usepackage[scaled=.92]{helvet} % for proper fonts
\usepackage{graphicx} % for EPS use the graphics package instead
\usepackage{balance}  % for useful for balancing the last columns
\usepackage{booktabs} % for pretty table rules
\usepackage{ccicons}  % for Creative Commons citation icons
\usepackage{ragged2e} % for tighter hyphenation

\title{Not a Technology Person: Motivating Older Adults Toward the Use of Mobile Technology}

\numberofauthors{6}
\author{%
  \alignauthor{%
    \textbf{Gabriela Villalobos-Z\'{u}\~{n}iga}\\
    \affaddr{University of Lausanne} \\
    \affaddr{gabriela.villalobos@unil.ch} }\alignauthor{%
    \textbf{Mauro Cherubini}\\
    \affaddr{University of Lausanne}\\
    \email{mauro.cherubini@unil.ch} }  }

\def\plaintitle{SIGCHI Extended Abstracts Sample File: Note Initial
  Caps} \def\plainauthor{First Author, Second Author, Third Author,
  Fourth Author, Fifth Author, Sixth Author}
\def\plainkeywords{Authors' choice; of terms; separated; by
  semicolons; include commas, within terms only; required.}

\hypersetup{%
  pdftitle={\plaintitle}, pdfauthor={\plainauthor},
  pdfkeywords={\plainkeywords}, }

\begin{document}

\maketitle

% Uncomment to disable hyphenation (not recommended)
% https://twitter.com/anjirokhan/status/546046683331973120
\RaggedRight{} 

% Do not change the page size or page settings.
\begin{abstract}
Older users population is rapidly increasing all over the World. Presently, we observe efforts in the human-computer interaction domain aiming to improve life quality of age 65 and over through the use of mobile apps. Nonetheless, these efforts focus primary on interface and interaction design. Little work has focused on the study of motivation to use and adherence to, of elderly to technology. Developing specific design guidelines for this population is relevant, however
it should be parallel to the study of desire of elderly to embrace specific technology in their life. Designers should not be limited to technology design but consider as well how to fully convey the value that technology can bring to the lives of the users and motivate adoption. This position paper discusses techniques that might nudge elderly towards the use of new technology.
\end{abstract}



\section{Introduction}

By 2050, population over age 65 will be almost 30\% in Europe and about 17\% in emergent economies \cite{fisk2009designing}. Older users population tend not to be organically attracted to technology; statistics from 2012 show that, 70\% of adults with ages over 65 or older own some type of mobile phone and 53\% expressed using the Internet. Moreover, with an age increase, the Internet reported use decreases to 34\% for adults over 75 years old \cite{zickuhr2012older}. This decline can be attributed to the lack of self-efficacy or absence of interest in technology \cite{vroman2015over,tsai2015getting}. However, population ages and the use of technology becomes relevant to the improvement of life quality. Technology in many of its forms has the potential to increase autonomy and reduce isolation \cite{tsai2015getting}, however if these benefits are not fully conveyed, we may experience low technology adoption levels. Nudging older users to adopt technology specifically designed to improve their lives becomes a determinant factor for the success of technological interventions. Much of previous research in this area focuses on feature design and testing (see for instance \cite{ijsselsteijn2007digital,haikio2007touch,sarcar2016towards}), however little work to date has focused on how to motivate this population to adopt technology. This position paper will focus specifically on this point. We will argue that by determining and understanding the motivation factors of older adults towards mobile technology usage, we can nudge them to value the self-improvement benefits it provides and ultimately adopting technological solutions.

\section{Potential Benefits}
The advantages for older users of using Internet-connected mobile devices are multiples. Features of apps and devices can be exclusively designed for their needs and abilities. For instance research suggest that the utilization of tablets, develop the feeling in older adults of being connected to the World and to their families, in addition to feeling more current and able to keep with trends \cite{tsai2015getting}. Also, systems designed for elderly can develop engagement into the desired activity. As another example, research reports that Flowie (a persuasive virtual coach) stimulated older adults to walk by showing them the amount of steps they had taken during the day \cite{albaina2009flowie}. Furthermore, mobile apps such as Oscarsenior, empowers older adults to maintain an independent life as well \cite{Oscarsenior}. Aging populations can benefit also from robot solutions that provide guidance for them in their environments and reminders to perform routine activities such as eating, drinking and taking medicines \cite{pollack2002pearl}. Unfortunately, identifying advantages of using technology as in the examples described above is often not obvious to users who have lived the majority of their lives without these solutions. Next, we will review some barriers to adoption identified in prior literature.

\section{Technology Adoption Barriers}
Despite stereotypes indicating that elderly are not well suited or interested in technology usage, research shows that they indeed perceive the benefits of its use to outweigh the cost of such use \cite{mitzner2010older}. \textit{Inconvenience} (e.g., unwanted calls, connection costs, mental effort to use mobile devices, discomfort of carrying the device all day, etc.), \textit{complexity of features design} (e.g., camera and pictures management on mobile phones, numbers of options and settings on mobile devices, etc.) and \textit{security and reliability} (e.g., lack of trust with the use of personal data, positioning technology not functioning when in the need, etc.) are considered the major dislikes from older adults to technology \cite{mitzner2010older}. In addition, other studies, suggests two main barriers to technology adoption: 1. \textit{Low computer self-efficacy}, 2. \textit{Performance anxiety connected with computer use} \cite{czaja2006factors}. Finally in addition to the points above, Tsai et al. \cite{tsai2015getting} suggest \textit{ergonomic impediments} as a barrier for technology adoption by this population. These barriers have similarities with the ones observed in younger adults (e.g. quality/quantity of the data provided by the service, interaction design of the service not corresponding to user needs) \cite{cherubini2011barriers}.
In this position paper, we argue that many of the barriers described above could be tackled by applying Persuasive Computing design techniques to the design of apps for mobile devices.

\section{Persuasive Design Techniques to Nudge Towards Adoption}
The Self-Determination Theory (SDT) \cite{deci2010self} propose three dimensions that describe human motivation, namely: \textit{autonomy} (e.g., people are in power of executing their own decisions), \textit{competence} (e.g., people are skillful to accomplish the task) and \textit{relatedness} (e.g., people feel important and connected to the main characters involved in the task they perform). When these three aspects are satisfied, a higher motivation level is reached, which in turns leads to technology adoption. In order to increase autonomy, competence and relatedness we turn to Persuasive Computing. 

Persuasive computing (PT) involves the use of a computing system or application which is intentionally designed to change a person's attitudes or behavior in an specific way \cite{fogg1999persuasive}.

In this position paper we argue that persuasive design can inform the conception of systems to aid older adults improve their social abilities at the same time that they observe an improvement in their well-being. We believe this can be achieved taking as a foundation the Persuasive System Design Framework (PSD) \cite{Oinas-kukkonen2009} and Self-determination theory of human motivation.

For the purpose of this workshop we will base our examples in the following PSD principles: \textit{tunneling} (persuading users while they are in the process of performing a task), \textit{reminders} (system reminds the users of their target behavior) and \textit{cooperation} (system fosters collaboration between their users). In the next section we will build a scenario to illustrate these three principles at play.

\section{Nudging Through Persuasive Design}

Mike is an older adult with basic technology skills: he uses his phone to search for information. He is not aware of the full set of features the phone has to offer. In this scenario, Mike wants to visit a park, he opens the browser and types "Sunshine Park" (see image 2). His purpose is to see the map and determine the path he needs to follow to arrive to his destination. As the system detects the intent, a tutorial is prompted on how to use the mobile for getting to the place (see image 5). 

In this initial part of the scenario, we can observe the \textit{tunneling} principle at play: the system provides tutoring while the user executes its primary task, in this case searching for a location. This aims at increasing their \textit{autonomy} in two ways: first, they are in no need of help to walk to the park; second, they learn a new feature of the phone without any support from other people. This approach may as well address the technological barrier of complexity of features design, by putting in place simple system actions (i.e., providing a contextual recommendation).

In the subsequent days while Mike is at home he suddenly hears his phone beeping (see image 10). The phone delivers a notification suggesting to explore a new location (see image 11). Mike accepts the suggestion. This time, he decides to visit a Museum.

By sending Mike a \textit{reminder} of the last time he visited the Park, and how he used the phone to find his way, the system continues to teach him the navigation functionality. This has the advantage of providing support to Mike's memory and in turn it enhances his level of \textit{competence}. Likely, next time he will be in the need of visiting an unknown place he will remember the turn-by-turn navigation functionality. This approach might addresses as well the adoption barrier related to of low computer self-efficacy. as older users will be able to increase their knowledge without any external help. 

Next on the scenario, Mike arrives at the Museum and meets a couple of older adults that struggled to find the place (see image 14). Mike decides to teach them how he used the turn-by-turn navigation to arrive safely to his destination (see image 15). Mike is able to empathize with them as he faced the same kind of frustration. Later that day, the phone suggests Mike if he would like to be friends with Tim and Alice, the couple he met earlier that day (see image 16). This happens as the system has detected the social interaction happening in a social context and as the user profiles are quite similar, it suggests to both parties whether they are interested in exchanging contacts. Only if both sides agree the information is shared. A week later, Tim and Alice want to visit the city Hall, but they are struggling to find the place. Since they remember Mikes suggestion to use the navigation system on their phone and felt they can trust him, they ask for his help to use the application. Tim and Alice did not feel ashamed to reveal their lack of knowledge because Mike has a similar life experience as themselves. 

In this last point, we can see how a persuasive system can develop \textit{collaboration} by peer to peer tutoring. Users exchange ideas and learn how to take advantage of the system features in a climate of trust and empathy. This develops the feeling of \textit{relatedness} with technology, as older adults can get acquainted to it through peers who share similar life backgrounds and knowledge \cite{de2010movipill}.

\section{Contribution and Goals in Participating to the Workshop }
We believe that persuasion has the potential to act as a key player  in motivating  and  developing a sense of interest in technology for aging  population. It is through persuasion that we aim to develop a positive behavior change that improves older adults well-being. However, this can not be executed in isolation  without considering aspects  of user-centered design, engineering and  health for instance. Likewise, new challenges are  approaching fast,  we can also  question ourselves, how to sustain behavior change over time? Or how to maintain older adults' motivation high over time even when the persuasive application is no longer at hand? These are some general questions we would like to discuss during the workshop.

More specifically, we are interested in getting feedback on our proposed scenario. For instance we would like to examine when persuasive  principles might break, or being detrimental to the user experience. Furthermore, we are interested in discussing the trade-offs between usability and persuasive design as some of the proposed techniques might further complicate the user experience. Also, we would love to discuss at which level of the 'development stack' these principles should be embedded, that being either the device level, or the OS level, or the application level. We do not have a unique answer to these points.

\balance{} 

\bibliographystyle{SIGCHI-Reference-Format}
\bibliography{sample}

\end{document}